\documentclass[preprint,3p,12pt]{elsarticle}

\usepackage{amssymb}
\usepackage{amsmath}
\newcommand{\ud}{\mathrm{d}}

\newcommand{\intl}{\int_{-\infty}^{+\infty}}

\DeclareMathOperator{\sgn}{sgn}

\journal{Physics Letters A}

\begin{document}

\begin{frontmatter}

\title{Short pulse equations and localized structures in frequency band gaps of nonlinear metamaterials}

\author[semfe]{N.L. Tsitsas}

\author[uoi]{T.P. Horikis}

\author[umass]{Y. Shen}

\author[umass]{P.G. Kevrekidis}

\author[umass]{N. Whitaker}

\author[uoa]{D.J. Frantzeskakis}

\address[semfe]{School of Applied Mathematical and Physical Sciences, National
Technical University of Athens, Zografos, Athens 15773, Greece}
\address[uoi]{Department of Mathematics, University of Ioannina, Ioannina 45110, Greece}
\address[umass]{Department of Mathematics and Statistics, University of Massachusetts,
Amherst MA 01003-4515, USA}
\address[uoa]{Department of Physics, University of Athens, Panepistimiopolis, Zografos, Athens 157 84, Greece}
\begin{abstract}
We consider short pulse propagation in nonlinear metamaterials characterized by a
weak Kerr-type nonlinearity in their dielectric response. In the frequency ``band
gaps" (where linear electromagnetic waves are evanescent) with linear effective
permittivity $\epsilon<0$ and permeability $\mu>0$, we derive two short-pulse
equations (SPEs) for the high- and low-frequency band gaps. The structure of the
solutions of the SPEs is also briefly discussed, and connections with the soliton
solutions of the nonlinear Schr\"odinger equation are presented.
\end{abstract}

\end{frontmatter}

Artificially engineered {\it metamaterials} demonstrate unique electromagnetic (EM)
properties \cite{reviews}. The frequency dependence of the effective permittivity
$\epsilon$ and permeability $\mu$ of these media is such that there exist frequency
bands where the medium displays either a right-handed (RH) behavior ($\epsilon>0$,
$\mu>0$) or a left-handed (LH) behavior ($\epsilon<0$, $\mu<0$), thus exhibiting
negative refraction at microwave \cite{exp1,exp2a,exp2b} or optical frequencies
\cite{shalaev}. Frequency band gaps, i.e., frequency domains where linear EM waves
are evanescent (e.g., for $\epsilon<0$ and $\mu>0$), also exist in metamaterials.
Hence, when a nonlinearity occurs, say in the dielectric response of the medium (a
physically relevant situation in {\it nonlinear metamaterials}
\cite{zharov,agranovich,shadrivov,lazarides-tsironis,scalora,wen-pra,yskss}), then
nonlinearity-induced localization of EM waves is possible. Such localization is
indicated by the formation of {\it gap solitons}, which occur mainly in nonlinear
optics \cite{aceves} and Bose-Einstein condensates (BECs) \cite{ourbook}, by means
of the nonlinear Schr\"{o}dinger (NLS) equation with a periodic potential. Gap
solitons were also predicted to occur in nonlinear metamaterials \cite{longhi}.
There, the approximation of slowly varying electric and magnetic field envelopes,
led to a nonlinear Klein-Gordon (NKG) equation supporting gap solitons.

Nonlinear models describing localization of wave packets in periodic media, e.g.,
the NLS equation in optics \cite{aceves} and NKG equation in metamaterials
\cite{longhi} are usually derived in the framework of the {\it slowly-varying
envelope approximation}. However, as far as {\it ultra-short} pulse propagation is
concerned, i.e, for pulse widths of the order of a few cycles of the carrier
frequency, the NLS or the NKG models may fail. Indeed, in the context of nonlinear
fiber optics, the proper model describing the evolution of such ultra-short pulses
was shown to be the so-called {\it short-pulse equation} (SPE) \cite{sw}. Numerical
simulations in Ref. \cite{sw} comparing solutions of Maxwell's equations to the
ones of the SPE and NLS models have shown that the accuracy of the SPE (NLS)
increases (decreases) as the pulse width shortens. More recently, motivated by the
fact that the SPE model has no smooth pulse solutions propagating with fixed shape
and speed, a {\it regularized} SPE (RSPE) was derived \cite{jones}. The RSPE
supports, under rather strict conditions, smooth traveling wave solutions.

In this work, we derive nonlinear evolution equations describing {\it ultra-short}
pulses that can be formed in the band gaps of nonlinear metamaterials. In doing so,
we consider a metamaterial characterized by the permittivity $\epsilon$ and
permeability $\mu$ of Ref. \cite{exp1}, as well as a weak Kerr-type nonlinearity in
the medium's dielectric response \cite{scalora,wen-pra,longhi}. Following the lines
of Refs. \cite{sw,jones}, we derive appropriate expressions for $\mu$ in the
high-frequency (HF) and low-frequency (LF) band gaps with $\epsilon<0$ and $\mu>0$.
Then, we use a multiscale perturbation method, with different small parameters for
the HF and LF gaps depending on the metamaterial characteristics, to derive from
Maxwell's equations two SPEs. Each of these equations describes the evolution of
ultra-short pulses either in the HF or the LF gap and can not be valid
simultaneously for a given metamaterial. We also discuss the structure of the
resulting peakon-like solutions of these equations, and draw parallels to NLS-like
soliton solutions (which can be regarded as ultra-short gap solitons in nonlinear
metamaterials). Such peakon-like and breather-like solutions are also briefly
considered numerically both in the context of the SPE models and in that of the
originating Maxwell's equation dynamics.

\begin{figure}[htbp]
\centering
\includegraphics[width=3in]{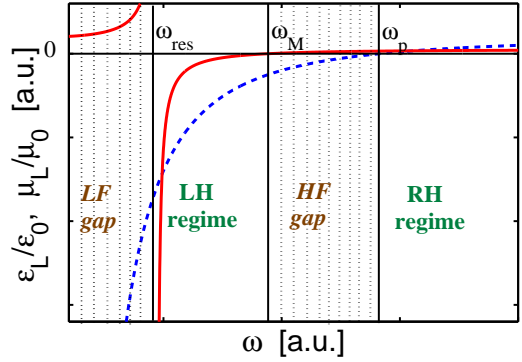}
\caption{
(Color online) Linear parts of the relative magnetic permeability, $\hat{\mu}/\mu_0$ [solid (red) line],
and electric permittivity, $\hat{\epsilon}/\epsilon_0$ [dashed (blue) line] as functions of the frequency $\omega$
in arbitrary units (a.u.). Propagation of linear waves is allowed ($\hat{\epsilon} \hat{\mu} >0$) for
$\omega>\omega_p$ and $\omega_{\rm res} < \omega < \omega_M$, where the medium displays right-handed (RH)
and left-handed (LH) behavior, respectively. Dotted regions with $\hat{\epsilon}<0$ and $\hat{\mu}>0$, namely,
$0 < \omega < \omega_{\rm res}$ (LF gap) and $\omega_M < \omega < \omega_p$ (HF gap), are frequency band
gaps where linear waves are evanescent.
}
\label{fig1}
\end{figure}

We consider a composite metamaterial consisting of an array of conducting wires and
split-ring resonators (SRRs), characterized by a frequency dependent effective
permittivity $\hat{\epsilon}(\omega)$ and magnetic permeability $\hat{\mu}(\omega)$
(below we will use $f$ and $\hat{f}$ to denote any function $f$ in the time- and
frequency-domain,  respectively). Here, the functions $\hat{\epsilon}(\omega)$ and
$\hat{\mu}(\omega)$ are assumed to take the form (in the lossless case)
\cite{exp1},
\begin{eqnarray}
\hat{\epsilon}(\omega)=\epsilon_{0}\bigg(1-\frac{\omega_p^2}{\omega^2}\bigg),
\quad
\hat{\mu}(\omega)=\mu_{0}\bigg(1-\frac{F\omega^2}{\omega^2-\omega_{\rm res}^2}\bigg),
\label{eq:e-m2}
\end{eqnarray}
where $\epsilon_0$ and $\mu_0$ are the vacuum permittivity and permeability,
$\omega$ is the frequency of the EM wave, $\omega_p$ the plasma frequency, \emph{F}
the filling factor, and $\omega_{\rm res}$ the resonant SRR frequency. Assuming
that $\omega_p > \omega_{\rm res}$, the dispersion relation $k^2 = \omega^2
\hat{\epsilon}(\omega) \hat{\mu}(\omega)$ ($k$ is the wavenumber) shows that linear
waves exist ($k^2 >0$) in the bands $\omega>\omega_p$ and $\omega_{\rm res} <
\omega < \omega_M \equiv \omega_{\rm res} /\sqrt{1-F}$, where the medium displays
right-handed (RH) and left-handed (LH) behavior, respectively. On the other hand,
there exist two frequency bands, with $\hat{\epsilon}<0$ and $\hat{\mu}>0$, where
linear EM waves are evanescent ($k^2<0$). These frequency ``band-gaps'' are $0 <
\omega < \omega_{\rm res}$ [``low-frequency'' (LF) gap], and $\omega_M < \omega <
\omega_p$ [``high-frequency'' (HF) gap]. Typical dispersion curves, showing
$\hat{\epsilon}(\omega)$ and $\hat{\mu}(\omega)$, are depicted in Fig. \ref{fig1}.

We now consider a nonlinear metamaterial, which can be realized by filling the
slits of the SRRs with a weakly nonlinear dielectric
\cite{zharov,shadrivov,lazarides-tsironis,scalora,yskss}. In particular, we assume
that this metamaterial exhibits a weak cubic (Kerr-type) nonlinearity in its
dielectric response \cite{scalora,wen-pra,longhi}, described by a nonlinear
polarization vector $\mathbf{P}_{NL}$ of the form,
\begin{eqnarray}
\mathbf{P}_{NL}=\epsilon_{0} \int_{-\infty}^{+\infty} \chi_{NL}(t-\tau_1, t-\tau_2, t-\tau_3)
(\mathbf{E}(\tau_1)\cdot \mathbf{E}(\tau_2)) \mathbf{E}(\tau_3) d\tau_1 d\tau_2
d\tau_3, \label{pnl}
\end{eqnarray}
where $\mathbf{E}$ is the electric field, and $\chi_{NL}$ is the nonlinear electric
susceptibility of the medium. In the case of small-amplitude, {\it ultra-short}
pulse propagation, the nonlinear response is  {\it instantaneous}, namely,
\begin{equation}
\chi_{NL}(t-\tau_1, t-\tau_2, t-\tau_3)
= \kappa \delta(t-\tau_1)\delta(t-\tau_2)\delta(t-\tau_3),
\label{xnl}
\end{equation}
where $\kappa$ is the Kerr coefficient given by $\kappa = \pm E_{c}^{-2}$, with
$E_c$ being a characteristic electric field value (e.g., of the order of $200$ V/cm
for n-InSb \cite{shadrivov}); generally, both cases of focusing ($\kappa>0$) and
defocusing ($\kappa<0$) dielectrics are possible. Notice that Eqs. (\ref{pnl}) and
(\ref{xnl}) imply that $\mathbf{P}_{NL}=\epsilon_0 \kappa (\mathbf{E} \cdot
\mathbf{E})\mathbf{E}$.

We now assume propagation along the $+z$ direction of a $x$- ($y$-) polarized
electric (magnetic) field, namely, $\mathbf{E}(z,t)=\hat{\mathbf{x}} E(z,t)$ and
$\mathbf{H}(z,t)=\hat{\mathbf{y}}H(z,t)$, where $\hat{\mathbf{x}}$ and
$\hat{\mathbf{y}}$ are the unit vectors in the $x$ and $y$ directions,
respectively. Then, Maxwell's equations lead to the following nonlinear wave
equation for $E(z,t)$:
\begin{equation}
\partial_z^2 E-\partial_t^2(\epsilon\ast\mu \ast E)-\epsilon_0 \kappa \partial_t^2 (\mu \ast E^3)=0,
\label{wee}
\end{equation}
where $\ast$ denotes the convolution integral, $f(t)\ast
g(t)=\int_{-\infty}^{+\infty}f(\tau)g(t-\tau)d\tau$, of any functions $f(t)$ and
$g(t)$. Note that once the electric field $E(z,t)$ is obtained from Eq.
(\ref{wee}), the magnetic field $H(z,t)$ can be derived from Faraday's law.

\begin{figure}[htbp]
\centering
\includegraphics[width=3in]{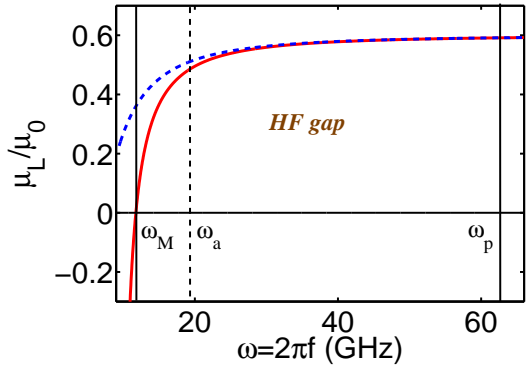}
\caption{
(Color online) Linear part of the relative permeability $\hat{\mu}/\mu_0$
in the HF gap ($F=0.4$, $\omega_{\rm res} = 2\pi \times 1.45$ GHz,
$\omega_M = 2\pi \times 1.87$ GHz, and $\omega_p = 2\pi \times 10$ GHz).
Solid (red) and dashed (blue) lines correspond, respectively, to the exact
[Eq. (\ref{eq:e-m2})] and approximate [Eq. (\ref{muap1})]
expressions of $\hat{\mu}(\omega)/\mu_0$ in this band. The approximation
produces a relative error less than $5\%$ for $\omega_{\rm a} \equiv 2\pi \times 3.1$
GHz $< \omega < \omega_p =2\pi \times 10$ GHz.
}
\label{fig2}
\end{figure}

We first consider the high-frequency (HF) band gap, $\omega_M < \omega < \omega_p$,
and assume EM wave frequencies $\omega$ such that $\omega \gg \omega_{\rm res}$. In
this regime, $\hat{\mu}(\omega)$ in Eq. (\ref{eq:e-m2}) is approximated by:
\begin{equation}
\hat{\mu}(\omega) \approx \mu_0(1-F) -\mu_0 F
\frac{\omega_{\rm res}^2}{\omega^2}.
\label{muap1}
\end{equation}
Using the physically relevant parameter values \cite{lazarides-tsironis} $F=0.4$,
$\omega_{\rm res} = 2\pi \times 1.45$ GHz, $\omega_M = 2\pi \times 1.87$ GHz, and
$\omega_p = 2\pi \times 10$ GHz , in Fig. \ref{fig2} we show the exact [Eq.
(\ref{eq:e-m2})] and approximate [Eq. (\ref{muap1})] expressions for the effective
permeability in this band. As seen, the above approximation produces a relative
error from the exact form of $\hat{\mu}(\omega)/\mu_0$ less than $5\%$ in a wide
sub-interval of frequencies in this band, namely for $\omega_{\rm a} \equiv 2\pi
\times 3.1$ GHz $< \omega < \omega_p =2\pi \times 10$ GHz.

The expression of Eq. (\ref{muap1}) is useful for simplifying terms of Eq.
(\ref{wee}) involving convolution integrals. In particular, the terms
$\epsilon\ast\mu \ast E$ and $\mu \ast E^3$ can respectively be approximated in the
frequency domain as $(1/c^2)(1-F)\hat{E}-(1/\omega^2 c^2)[(1-F)\omega_p^2 +
F\omega_{\rm res}^2]\hat{E}$ and $(1/c^2)\kappa (1-F)\widehat{E^3}-(1/\omega^2
c^2)\kappa F\omega_{\rm res}^2 \widehat{E^3}$. Here, $\hat{E}=\intl E\,\exp(i\omega
t)\,\ud t$ is the Fourier transform of $E$ and $c$ the velocity of light in vacuum.
As a result, Eq. (\ref{wee}) can be reduced to the form:
\begin{eqnarray}
\partial_z^2 E-\frac{1-F}{c^2}\partial_t^2 E - \frac{1}{c^2} [F\omega_{\rm res}^2+(1-F)\omega_p^2] E
- \frac{\kappa}{c^2}\left[ F \omega_{\rm res}^2 E^3 + (1-F)\partial_t^2 E^3
\right]=0. \label{wehf}
\end{eqnarray}
Next, measuring time, space, and the field intensity $E^2$ in units of $\omega_{\rm
res}^{-1}$, $v/\omega_{\rm res}$ [where $v = c(1-F)^{-1/2}$], and $|\kappa|^{-1}$,
respectively, Eq. (\ref{wehf}) is expressed in dimensionless form as
\begin{eqnarray}
\big( \partial_z^2 - \partial_t^2  - \tilde{\alpha} \big) E = s\left( \frac{F}{1-F} + \partial_t^2 \right) E^3,
\label{wehfn}
\end{eqnarray}
where $s=\sgn(\kappa)=\pm 1$ for focusing or defocusing nonlinearity, respectively,
and $\tilde{\alpha} = F/(1-F) + (\omega_p/\omega_{\rm res})^2$. We assume that
$(\omega_p/\omega_{\rm res})^2 \gg 1$ (as, e.g., in Ref. \cite{lazarides-tsironis})
in order  to ensure the validity of the approximation (\ref{muap1}) in a wide
sub-interval of the HF band gap (see above). Hence, $\tilde{\alpha}$ is a large
parameter, which suggests that $\tilde{\alpha} = \alpha/\varepsilon$, where
$\varepsilon$ is a formal small parameter (which sets also the field amplitude as
per our perturbative approach below), and $\alpha =\mathcal{O}(1)$. Furthermore,
considering propagation of small-amplitude short pulses, we introduce a multiple
scale ansatz of the form
\begin{equation}
E= \varepsilon^{3/2} E_1 (T_{\rm HF}, Z_1, \cdots)
+\varepsilon^{5/2} E_2 (T_{\rm HF}, Z_1, \cdots)+\cdots,
\label{an}
\end{equation}
where $T_{\rm HF}= \varepsilon^{-2}(t-z)$ and $Z_n= \varepsilon^n z$
($n=1,2,\cdots$). Substituting Eq. (\ref{an}) into Eq. (\ref{wehfn}), we obtain
various equations at different orders of $\varepsilon$. In particular, terms at
$\mathcal{O}(\varepsilon^{-5/2})$ cancel, there are no terms at
$\mathcal{O}(\varepsilon^{-3/2})$, while terms at
$\mathcal{O}(\varepsilon^{-1/2})$, cancel provided that the field $E_1$ satisfies
the following equation,
\begin{eqnarray}
2 \partial_{\zeta} \partial_{T_{\rm HF}} E_1 + \alpha E_1 + s\partial_{T_{\rm HF}}^2 E_1^3 = 0,
\label{spe}
\end{eqnarray}
where we have used the notation $\zeta \equiv Z_1$. Equation (\ref{spe}) is the
so-called SPE, which was derived in Ref. \cite{sw} as an appropriate model
describing the propagation of ultra-short pulses in silica optical fibers with a
Kerr nonlinearity.

\begin{figure}[htbp]
\centering
\includegraphics[width=3in]{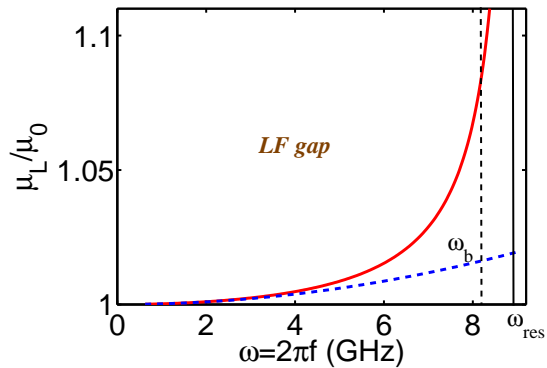}
\caption{
(Color online) Linear part of the relative permeability $\hat{\mu}/\mu_0$
in the LF gap ($F=0.02$ and $\omega_{\rm res} = 2\pi \times 1.45$ GHz).
Solid (red) and dashed (blue) lines correspond, respectively, to the exact [Eq. (\ref{eq:e-m2})],
and approximate [Eq. (\ref{muap2})] expressions of $\hat{\mu}(\omega)/\mu_0$ in this band.
The approximation produces a relative error less than $5\%$ for
$\omega_{\rm b} \equiv 2\pi \times 1.28$ GHz $< \omega < \omega_{\rm res}= 2\pi \times 1.45$ GHz.
}
\label{fig3}
\end{figure}

Next, consider the low-frequency (LF) band gap, $0 < \omega < \omega_{\rm res}$,
and assume that the EM frequency is $\omega \ll \omega_{\rm res}$. Then,
$\hat{\mu}(\omega)$ in Eq. (\ref{eq:e-m2}) is approximated by
\begin{equation}
\hat{\mu}(\omega) \approx \mu_0
\left(1 + F \frac{\omega^2}{\omega_{\rm res}^2}\right).
\label{muap2}
\end{equation}
Using $F=0.02$, $\omega_{\rm res} = 2\pi \times 1.45$ GHz, in Fig. \ref{fig3} we
show the exact [Eq. (\ref{eq:e-m2})] and approximate [Eq. (\ref{muap2})]
expressions for $\hat{\mu}(\omega)$ in the LF band gap. This produces a relative
error less than $5\%$ in a wide sub-interval of frequencies in this band, i.e., for
$0< \omega < \omega_{\rm b} \equiv 2\pi \times 1.28$ GHz.

We now employ Eq. (\ref{muap2}) to simplify the convolution integral terms of Eq.
(\ref{wee}). The terms $\epsilon\ast\mu \ast E$ and $\mu \ast E^3$ can respectively
be approximated in the frequency domain as $(1/c^2)[F \omega^2/\omega_{\rm res}^2 +
(1- F\omega_p^2 / \omega_{\rm res}^2)-\omega_p^2/\omega^2]\hat{E}$ and
$(1/c^2)\kappa (1+F \omega^2/\omega_{\rm res}^2) \widehat{E^3}$.
As a result, Eq. (\ref{wee}) reads,
\begin{eqnarray}
\partial_z^2 E-\frac{1}{c^2} \left(1-F \frac{\omega_p^2}{\omega_{\rm res}^2}\right) \partial_t^2 E -
\frac{\omega_p^2}{c^2}E =
- \frac{F}{\omega_{\rm res}^2 c^2}  \partial_t^4 E + \frac{\kappa}{c^2}\partial_t^2 E^3 -
\frac{\kappa F}{\omega_{\rm res}^2 c^2}\partial_t^4 E^3.
\label{welf}
\end{eqnarray}
Notice that in Eq. (\ref{welf}), the ratio $(\omega_p / \omega_{\rm res})^2$ is
considered to be a $\mathcal{O}(1)$ parameter (as, e.g., in Ref. \cite{longhi})
since $\omega_p$ is not involved in the band width of the LF band gap. In this
band, it is convenient to use a different small parameter, namely the filling
factor $F$, which is a physically relevant choice for SRRs
\cite{lowF_srr_model_a,lowF_srr_model_b}, as well as for other types of
metamaterials \cite{lowF_other_models_a,lowF_other_models_b,lowF_other_models_c}.
Then, measuring time, space, and the field intensity $E^2$ in units of $\omega_{\rm
res}^{-1}$, $c/\omega_{\rm res}$ and $|\kappa|^{-1}$, respectively, we reduce Eq.
(\ref{welf}) to the dimensionless form:
\begin{equation}
\bigg( \partial_z^2 - \partial_t^2  - \frac{\omega_p^2}{\omega_{\rm res}^2} \bigg) E
= s\partial_t^2 E^3.
\label{welfn}
\end{equation}
Next, we can again derive from Eq. (\ref{welfn}) a SPE for the LF band using
the asymptotic expansion
\begin{equation}
E= \varepsilon E_1 (T_{\rm LF}, Z_1, \cdots)
+\varepsilon^2 E_2 (T_{\rm LF}, Z_1, \cdots)+\cdots,
%\sum_{n=1}^{+\infty} \varepsilon^{n+\frac{1}{2}} E_n (\tau, z_1, z_2, \cdots),
\label{anlf}
\end{equation}
where $T_{\rm LF}= \varepsilon^{-1}(t-z)$ and $Z_n = \varepsilon^n z$
($n=1,2,\cdots$). Then, substituting Eq. (\ref{anlf}) into Eq. (\ref{welfn}), we
find that terms at $\mathcal{O}(\varepsilon^{-1})$ cancel, there are no terms at
$\mathcal{O}(\varepsilon^0)$, while terms at $\mathcal{O}(\varepsilon)$, cancel
provided that $E_1$ satisfies the following SPE,
\begin{eqnarray}
2 \partial_{\zeta} \partial_{T_{\rm LF}} E_1 + \frac{\omega_p^2}{\omega_{\rm res}^2} E_1
+ s\partial_{T_{\rm LF}}^2 E_1^3=0,
\label{rspe}
\end{eqnarray}
where again $\zeta \equiv Z_1$. In the above analysis, the filling factor $F$ was
treated as $\mathcal{O}(\varepsilon^j)$, with $j\ge 5$. However, if the filling
factor $F$ was assumed to be $\mathcal{O}(\varepsilon^4)$ then the additional term
$-\partial_{T_{\rm LF}}^4 E_1$ would appear in the left-hand side of Eq.
(\ref{rspe}). In such case, Eq. (\ref{rspe}) would then be the so-called
regularized SPE (RSPE) model, which was recently derived in Ref. \cite{jones}, also
in the context of ultra-short pulse propagation in nonlinear optical fibers.
However, due to its negative sign, the term $-\partial_{T_{\rm LF}}^4 E_1$ does not
have the regularizing effect of \cite{jones}, and higher-order regularizations,
outside the scope of the present work, may need to be considered.

Solutions of the SPEs are now briefly discussed. First, we unify SPEs of Eqs. (\ref{spe}) and
(\ref{rspe}) in the single equation
\begin{eqnarray}
\partial_{\zeta} \partial_{\tau} u + \gamma u
+ \frac{1}{2} s \partial_{\tau}^2 u^3=0,
\label{spe-rspe}
\end{eqnarray}
where $u \equiv E_1$, while $\gamma=\alpha/2$ and $\tau=T_{\rm HF}$, or
$\gamma=\omega_p^2/(2\omega_{\rm res}^2)$ and $\tau = T_{\rm LF}$ for the HF or the
LF band gap, respectively. Then, seeking traveling wave solutions of the form
$u=u(\xi)$, where $\xi=\zeta-C \tau$ (with $C$ being associated with the velocity
of the traveling wave), Eq. (\ref{spe-rspe}) is reduced to the following ordinary
differential equation:
\begin{equation}
-Cu_{\xi\xi}+\gamma u+\frac{1}{2}sC^2(6uu_{\xi}^2+3u^2u_{\xi\xi})=0.
\label{ode1}
\end{equation}
The transformation $u_{\xi}^2=w(u)$ produces a linear equation with respect to
$w(u)$ which can be solved to give
\begin{gather}
u_{\xi}^2=- \frac{\gamma u^2}{C}\frac{3sCu^2-4}{(3sCu^2-2)^2},
\label{ux}
\end{gather}
subject to the initial condition $u_\xi(\pm\infty)=u(\pm\infty)=0$ or
$w(u(\pm\infty))=w(0)=0$. The sign of the product $sC$ is rather important. Indeed,
if $sC<0$ then the solutions are always either ascending or descending since no
maxima or minima can occur. Thus, we are left with the choice $sC>0$, which allows
bounded solutions. In the case $s=+1$ (i.e., focusing dielectrics with $\kappa>0$),
which implies that $C>0$, the maximum of the traveling wave occurs when $u_\xi=0$
or, equivalently, $u=\sqrt{4/3C}$. To avoid the singularity at $u=\sqrt{2/3C}$ we
consider small amplitude pulses (essentially pulses that never reach the
singularity) and Eq. (\ref{ux}) is reduced to
the equation: $u_{\xi}^2=(\gamma/C)u^2$. The latter, possesses a {\it peakon}-like
solution (see, e.g., Ref. \cite{peakon}) of the form,
\begin{equation}
u(\xi)=A \exp(-\sqrt{\gamma/C}|\xi|),
\label{p}
\end{equation}
whose derivative has a discontinuity at $\xi =0$, with amplitude $A < \sqrt{2/3C}$.
Also, when the above approximation is not used, the right-hand side of Eq.
(\ref{ux}) needs to be positive giving a range of values for the field
$|u|<\sqrt{4/3|C|}$. These observations are consistent with the fact that the SPE
exhibits loop-solitons found in Ref. \cite{ss}.

\begin{figure}[htbp]
\begin{center}
\includegraphics[width=3in]{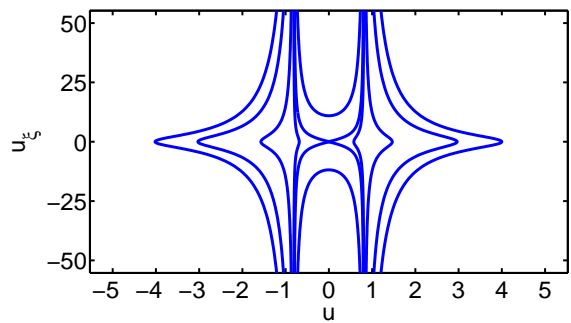}\caption{\label{phase}
Typical phase plane curves $(u,u_{\xi})$ associated with Eq.(\ref{ode1}) for $C=s=1$.}
\end{center}
\end{figure}

To further illustrate, we now conduct a phase plane analysis of Eq. (\ref{ode1}),
see Fig. \ref{phase} ($C=s=1$). From the relevant curves, it becomes clear that
there is no homoclinic orbit surrounding the fixed point at the origin. The only
possibility is that we move on one of the phase plane curves, say, in the upper
half plane up to a certain point, then ``jump'' from $(u,u_{\xi})$ to
$(u,-u_{\xi})$ and then, due to reversibility, return along the symmetric curve in
the lower half plane. This would constitute the peakon-like solutions discussed
above in the regime where $u$ is small. However, we should also note that when
integrating Eq. (\ref{spe-rspe}), we were not able to observe robust propagation of
such a waveform in the dynamics (hence it is not discussed further herein).

In addition, based on the formal connection between the SPE and the sine-Gordon
equation (SGE), a smooth approximate, sech-shaped, envelope soliton solution of the
SPE, based on the breather solution of the SGE, was derived in \cite{ss}. In the
framework of Eq. (\ref{spe-rspe}), this solution has the approximate form
\begin{eqnarray}
u \approx 4m (3\gamma)^{-1/2} \cos\left(\zeta + \gamma \tau \right)
{\rm sech}\left[ m \left(\zeta - \gamma \tau \right)\right],
\label{gs1}
\end{eqnarray}
where $m$ is an arbitrary real parameter, $0< m<1$. It is clear that the shape of
the SPE pulse in Eq. (\ref{gs1}) bears resemblance to the NLS soliton, as it
consists of a ${\rm sech}$-shaped pulse modulating a periodic function. The
existence of this approximate envelope soliton solution, characterized by a small
amplitude and inverse width (both determined by the parameter $m<1$), suggests a
connection between the SPE and the NLS equation. Such a connection can be
established using the method of multiple scales as follows: introducing the
variables $\zeta_n = \varepsilon^n \zeta$, $\tau_n = \varepsilon^n \tau$ (with
$\varepsilon \ll 1$ and $n=0,1,2,\cdots$) and expanding the field as $u=
\sum_{n=1}^{+\infty} \varepsilon^n u_n (\tau_n, \zeta_n)$, we obtain from Eq.
(\ref{spe-rspe}) (for $s=+1$) the following results. The unknown field $u_1$ is
found to be of the form
\begin{equation}
u_1=A(\zeta_1,...,\tau_1,...)\exp[i(k\zeta_0-\omega\tau_0)]+ {\rm c.c.}
\label{u1}
\end{equation}
where ${\rm c.c.}$ denotes complex conjugate, while the wavenumber $k$ and
frequency $\omega$ are connected by the dispersion relation [found to order
$\mathcal{O}(\varepsilon)$]: $\omega k+\gamma=0$. On the other hand [as found at
$\mathcal{O}(\varepsilon^3)$], the unknown envelope function $A$ satisfies the
following NLS equation,
\begin{eqnarray}
i\left(\partial_{\zeta_2}A+k'\partial_{\tau_2}A\right)
-\frac{k''}{2}\partial^2_{\bar{\tau}_1}A
+\frac{3}{2}\omega|A|^2A=0,
\label{NLS}
\end{eqnarray}
where $\bar{\tau}_1=\tau_1-k'\zeta_1$, $k'\equiv \partial k/\partial \omega
=-k/\omega$, and $k''\equiv \partial^2 k/\partial \omega^2 =2k/\omega^2$. Thus, it
is clear that the well-known sech-shaped envelope soliton solution of the NLS Eq.
(\ref{NLS}) resembles the soliton of Eq. (\ref{gs1}), and scales in space and time
in a similar way (if $m$ is of O$(\varepsilon)$). Such smooth solutions of the SPE
models derived above can be regarded as weak gap solitons (in the respective
scales) that can be formed in the HF and LF band gaps of the considered nonlinear
metamaterials.

\begin{figure}[htbp]
\centering
\includegraphics[width=2.2in]{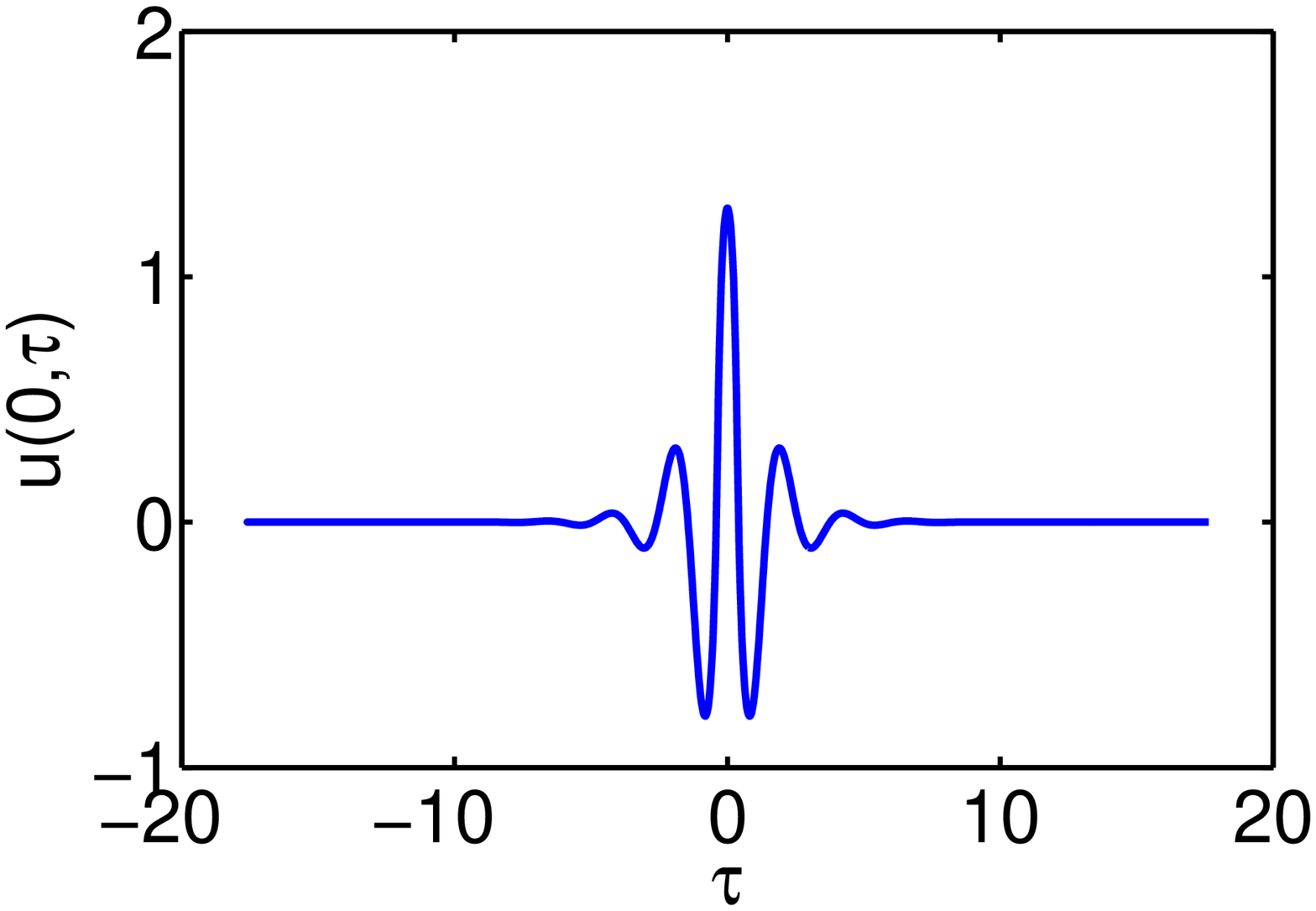}
\includegraphics[width=2.2in]{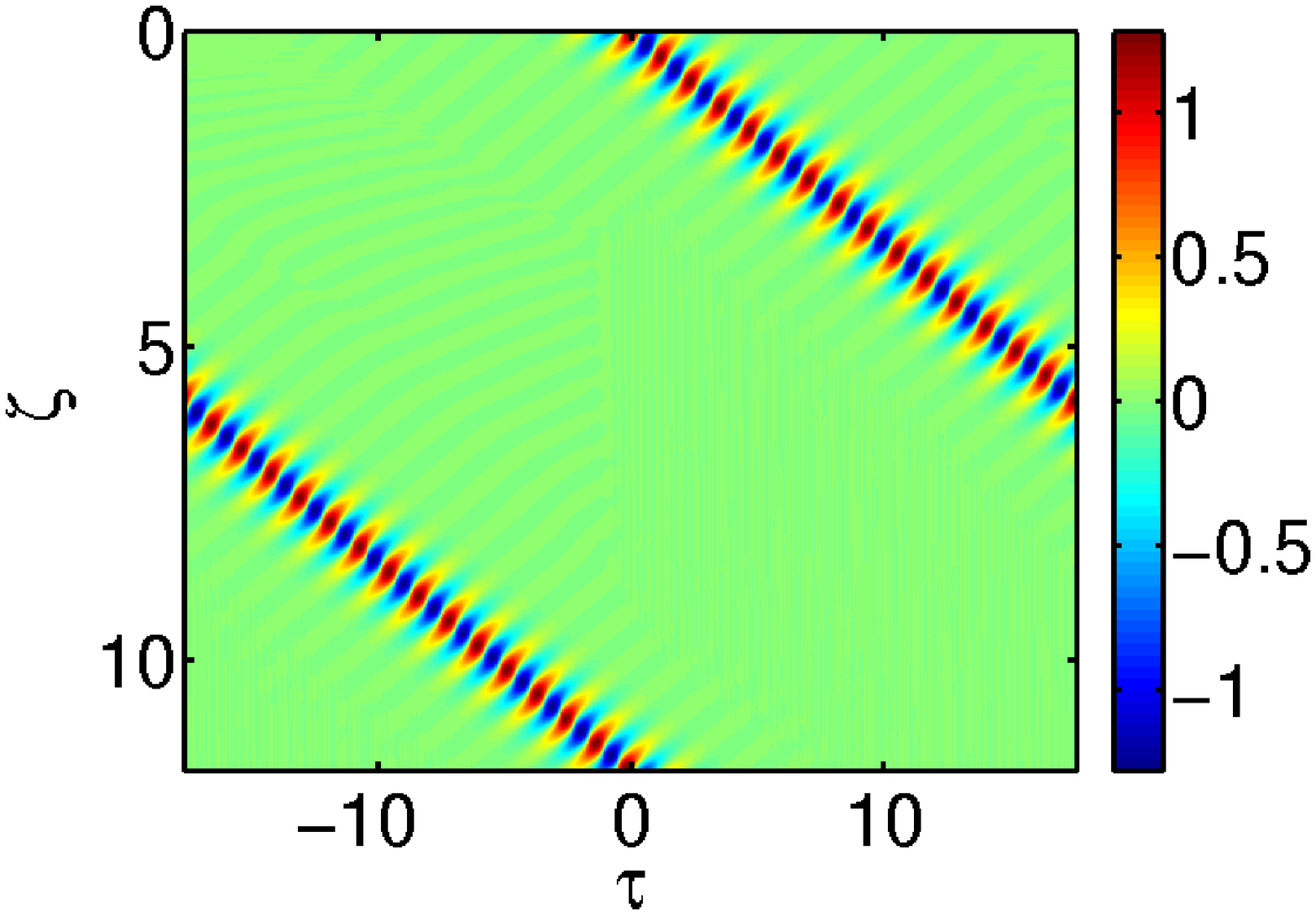}\\
\includegraphics[width=2.2in]{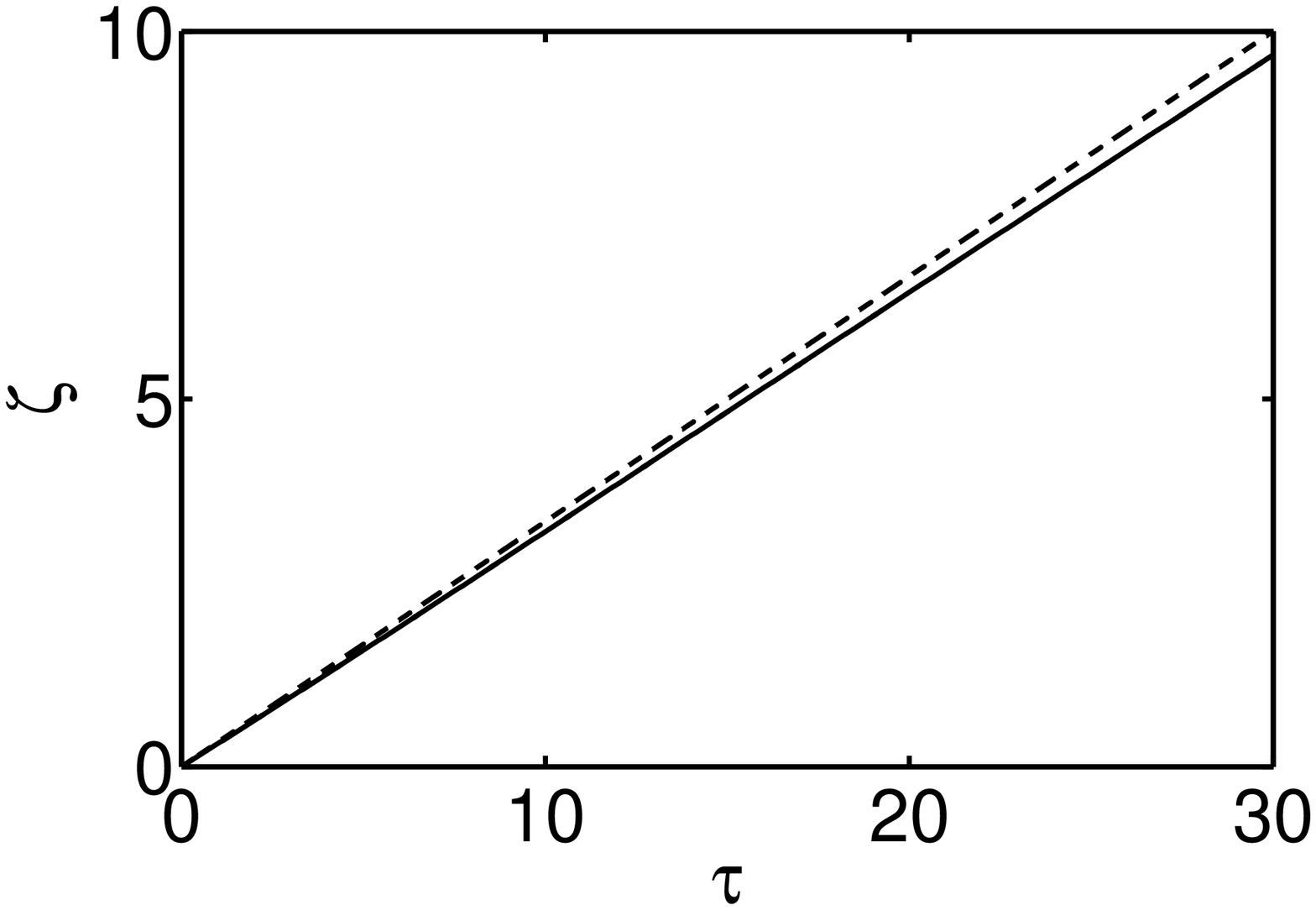}
\includegraphics[width=2.2in]{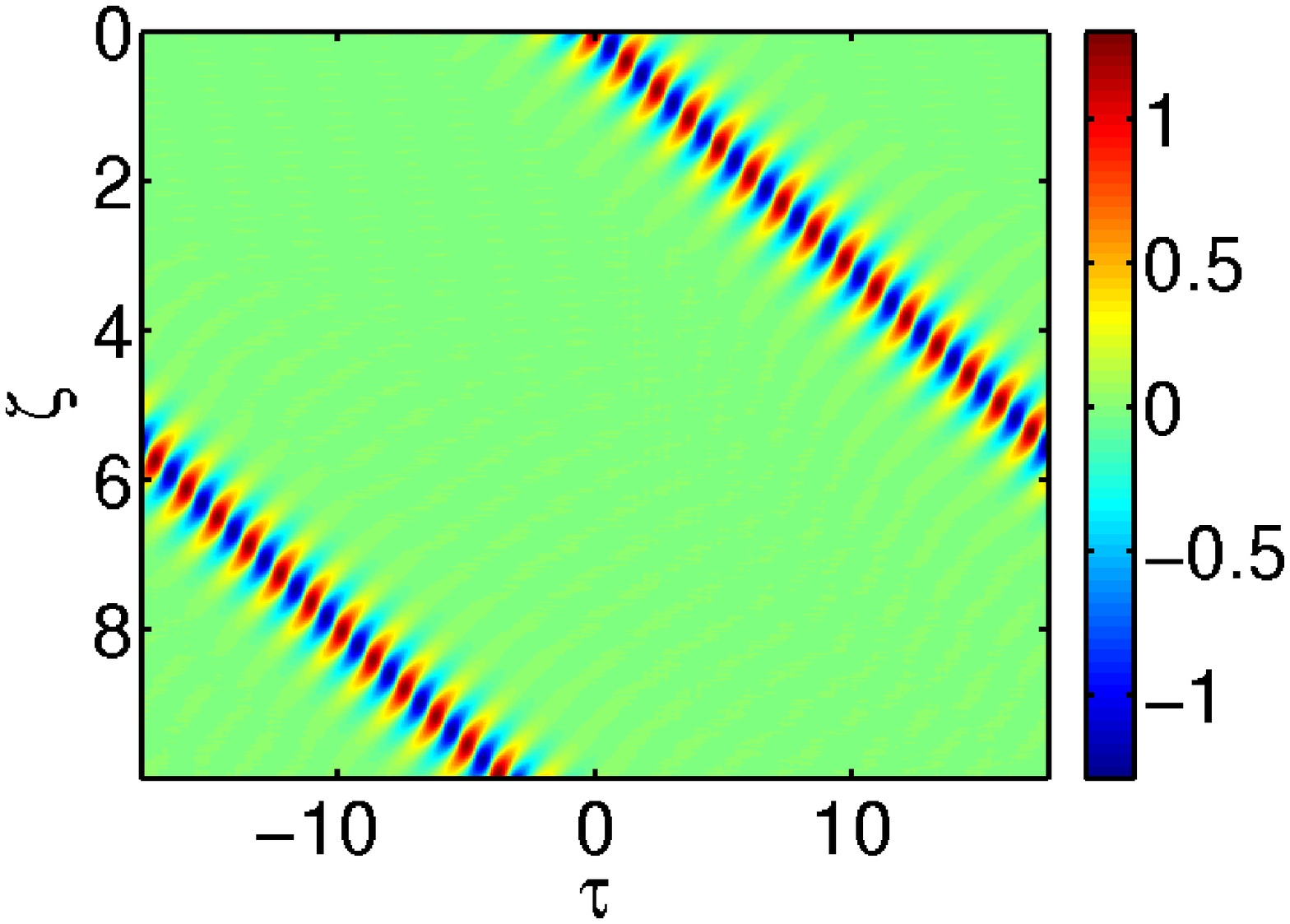}
\caption{The top left panel shows the breather initial condition used in
Eqs. (\ref{spe-rspe}) and (\ref{wehfn}). The bottom left panel shows the evolution
of the center of mass of the breathers in the
two respective models by dotted and solid lines. The top right panel
shows the space-time contour plot of the field evolution
with the breather initial condition in Eq. (\ref{spe-rspe}). The
bottom right panel shows the same for Eq. (\ref{wehfn}).}
\label{breather}
\end{figure}

To corroborate these results, we have performed numerical simulations of both Eq.
(\ref{spe-rspe}), as well as of Eq. (\ref{wehfn}) from which the former was
derived. Our numerical method relies on Fourier transforming Eq. (\ref{spe-rspe})
with respect to $\tau$, then solving the ensuing first order ODE in $\zeta$ (for
each frequency), via a fourth-order Runge-Kutta scheme, and then Fourier
transforming back to obtain $u(\zeta,\tau)$. In the simulations below, we use
$s=1$, $F=0.4$, and $\omega_p/\omega_\mathrm{res}=10/1.45$ and $\varepsilon=0.1$.
The initial condition is shown in the top left panel of Fig. \ref{breather} and is
obtained as the exact breather solution of the SPE equation (Eq. (22) in \cite{ss},
with $m = 0.32$). The evolution of the breather can be seen both from the bottom
left panel of Fig. \ref{breather}, showing the center of mass of the solution vs.
time and the top right panel illustrating the contour plot of the full space-time
evolution. This is clearly a robust localized structure which propagates through
the domain with constant speed in time (under the used periodic boundary
conditions). We have also integrated Eq. (\ref{wehfn}) with the same type of
breather-like initial profile. In the latter case, however, from the multiple scale
ansatz, the initial condition of Eq. (\ref{wehfn}) was chosen as
$E(0,\tau)=\varepsilon u(0,\tau), \ E_z(0,\tau) = -u_\tau(0,\tau)$. Furthermore,
here one needs to be careful, similarly to what was done in \cite{CJSW}, to
eliminate the propagation at very low frequencies [below $\sqrt{\tilde{\alpha}}$ in
the setting of Eq. (\ref{wehfn})]. The result of the time integration is shown in
the bottom right of Fig. \ref{breather} --see also the solid line showing the
center of mass evolution in the bottom left panel of the figure. It is clear that
the breather is robust in this setting as well, although its propagation speed is
slightly smaller than that of the SPE breather. This result confirms our prediction
that such ``gap breathers'' should be observable in nonlinear metamaterials of the
type considered in this work.

In conclusion, we derived short-pulse equations (SPEs) describing the propagation
of ultra-short pulses in nonlinear (Kerr-type) metamaterials. Two SPEs were found
for the high- and low-frequency band gaps, respectively, characterized by a
negative (positive) linear effective permittivity (permeability), where propagation
of linear electromagnetic (EM) waves is not allowed. We also discussed the
structure of the solutions of the SPEs and presented the approximate peakon-like
and breather-like solitary waves, which can be regarded as weak ultra-short gap
solitons. We also examined these structures via numerical computations to
illustrate the apparent non-robustness of the former, and stable propagation of the
latter. Generally, the existence of such structures, indicates the possibility of
nonlinear localization of EM waves in the gaps of nonlinear metamaterials.
Interesting subjects for future research would include systematic studies of the
stability of such ultra-short gap solitons both in the framework of the SPEs and
Maxwell's equations as well as higher dimensional generalizations of the structures
considered herein.

\section*{Acknowledgements}
The work of DJF was partially supported by the Special Account for Research Grants
of the University of Athens.

\section*{References}
%\bibliographystyle{elsarticle-num}
%\bibliography{spe_refs}

\end{document}